\documentclass[doublecol]{epl2}
\usepackage{amsmath}
\usepackage{amssymb}
\usepackage{bm}
\usepackage{graphicx}
\usepackage{color}

\title{Colored noise induces synchronization of limit cycle oscillators}
\shorttitle{Colored noise induces synchronization of limit cycle oscillators}

\author{Wataru Kurebayashi\inst{1} \and Kantaro Fujiwara\inst{1} \and
Tohru Ikeguchi\inst{1,2}}
\shortauthor{W. Kurebayashi \etal}

\institute{
  \inst{1} Graduate School of Science and Engineering, Saitama University
	- 255 Shimo-ohkubo, Sakura-ku, Saitama-city, Saitama, 338-8570 Japan \\
  \inst{2} Saitama University Brain Science Institute
	- 255 Shimo-ohkubo, Sakura-ku, Saitama-city, Saitama, 338-8570 Japan
}

\abstract{
Driven by various kinds of {noise},
ensembles of limit cycle oscillators can synchronize.
In this Letter, we propose a general formulation
of synchronization of the
oscillator ensembles driven by common colored noise
with an arbitrary power spectrum.
To explore statistical properties of {such 
colored noise-induced synchronization},
we derive the stationary distribution of the phase difference
between {two} oscillators {in the ensemble}.
{This analytical result theoretically predicts various synchronized
and clustered states induced by colored noise and also clarifies that
these phenomena have a different synchronization mechanism from the case of white noise.}
}
\begin{document}

\maketitle

\section{Introduction}

Driven by common {noise},
many nonlinear dynamical systems
can synchronize.
This phenomenon is called noise-induced synchronization,
which is observed in various 
kinds of the nonlinear dynamical systems,
for example, neural networks
\cite{mainen95__reliab_of_spike_timin_in_neocor_neuron,
galan06__correl_induc_synch_of_oscil},
electric circuits
\cite{yoshida06__noise_induc_synch_of_uncoup_nonlin_system},
electronic devices
\cite{utagawa08__noise_induc_synch_among_sub},
microbial cells
\cite{zhou05__molec_commun_throug_stoch_synch},
lasers
\cite{uchida04__consis_of_nonlin_system_respon}
{and chaotic dynamical systems
\cite{zhou,wang}}.
It has been theoretically proven that limit cycle oscillators
can synchronize driven by common noise
\cite{teramae04__robus_of_noise_induc_phase}.
Many studies
have investigated the synchronization
property in case of various types of drive noises,
for example, Gaussian white noise
\cite{nakao07__noise_induc_synch_and_clust,
yoshimura08__invar_of_frequen_differ_in,
nagai10__noise_induc_synch_of_large}
and Poisson impulses
\cite{nakao05__synch_of_limit_cycle_oscil}.
In ref. \cite{nakao07__noise_induc_synch_and_clust},
using a formulation of limit cycle oscillators driven by
common and independent Gaussian white noises,
Nakao et al. analytically obtain{ed}
the probability density function (PDF)
of phase differences between two oscillators,
which enables us to effectively characterize the synchronization property.
However, although there are some numerical studies{\cite{wang,
yoshimura07__synch_induc_by_common_color,
hata10__synch_of_uncoup_oscil_by}},
analytical conventional studies are limited to
the case that drive signals are white noise (temporally uncorrelated noise).
If we can assume that the drive signal is white noise,
we can use the Fokker-Planck approximation
\cite{risken96__fokker_planc_equat} {to explore statistical
properties of oscillator ensembles}.
However, such an ideal condition is rare in the real world.
For example, in neural circuits,
it is known that colored noise with negative autocorrelation
plays a key role to propagate synchronous activities
\cite{cateau06__relat_between_singl_neuron_and}.
However, it still remains unclear how the oscillators behave
if they are driven by common colored noise.

Recently, it has been clarified how a limit cycle oscillator
behaves if it is driven by colored non-Gaussian noise 
\cite{teramae06__noise_induc_phase_synch_of,
nakao10__effec_long_time_phase_dynam,
goldobin10__dynam_of_limit_cycle_oscil}.
In this Letter, {utilizing effective white-noise
Langevin description proposed
in ref. \cite{nakao10__effec_long_time_phase_dynam}},
we extend the formulation
in {r}ef. \cite{nakao07__noise_induc_synch_and_clust}
to colored noise
that has an arbitrary power spectrum.
We then analytically derive the PDF of the phase difference
between the oscillators if these oscillators
are driven by common colored noise.
We also conducted numerical simulations to
verify our analytical results.
The results show that
{the PDF of the phase difference
explicitly depends on the power spectrum of
the drive noise}.

\section{Model}

We used the following system
that consists of $N$ identical limit cycle oscillators
subject to common and independent multiplicative colored noises.
The dynamics of the $j$th oscillator
is described by
\begin{eqnarray}
 \dot{\bm{X}}^{(j)} &=& \bm{F}(\bm{X}^{(j)}) + \sqrt{D}
	\bm{G}(\bm{X}^{(j)})\bm{\xi}(t)
	\nonumber \\
	&&+ \sqrt{\epsilon}\bm{H}(\bm{X}^{(j)})\bm{\eta}^{(j)}(t),
	\label{eq. lim osci}
\end{eqnarray}
for $j=1,\ldots,N$,
where $\bm{X}^{(j)}\in\mathbb{R}^n$ is the
$n$-dimensional state variable of the $j$th oscillator;
$\bm{F}(\bm{X}^{(j)})\in\mathbb{R}^n$ is an unperturbed vector field
that has a stable $T$-periodic limit cycle orbit $\bm{S}(t)$;
$\bm{\xi}(t)\in\mathbb{R}^m$ is the common noise,
which drives all of the oscillators;
$\bm{\eta}^{(j)}(t)\in\mathbb{R}^m$
($j=1,\ldots,N$)
is the independent noise, which is
received independently by each oscillator;
$\bm{G}(\bm{X}^{(j)})\in\mathbb{R}^{n\times m}$ and
$\bm{H}(\bm{X}^{(j)})\in\mathbb{R}^{n\times m}$
represent how the oscillators
are coupled to the common and independent noises;
$D$ and $\epsilon$ are parameters to control
the intensities of the common and independent noises.
We introduced the following three assumptions:
(i) $\bm{\xi}(t)\in\mathbb{R}^m$
and $\bm{\eta}^{(j)}(t)\in\mathbb{R}^m$
are independent, identically distributed
zero-mean colored noises,
namely, $\langle \bm{\xi}(t) \rangle=\bm{0}$,
$\langle \bm{\eta}^{(j)}(t) \rangle=\bm{0}$,
$\langle \bm{\xi}(t)\bm{\eta}^{(j)}(s)^{\top} \rangle=\bm{O}$, and
$\langle \bm{\eta}^{(j)}(t)\bm{\eta}^{(k)}(s)^{\top} \rangle=\bm{O}$
($j\neq k$), where $\top$ denotes the transpose
{and $\langle\cdot\rangle$ represents
the temporal average};
(ii) $\bm{\xi}(t)$ and $\bm{\eta}^{(j)}(t)$
can be approximated as the convolution
of an arbitrary filter function and white noise;
and (iii) $\bm{\xi}(t)$ and $\bm{\eta}^{(j)}(t)$
have correlation times shorter than
the time scale of the phase diffusion
($\sim O(D^{-\frac{1}{2}},\epsilon^{-\frac{1}{2}})$).

To characterize the statistical properties
of the drive noises $\bm{\xi}(t)$ and $\bm{\eta}^{(j)}(t)$,
{we define correlation matrices
$\bm{C}_{\xi}(\tau)\in\mathbb{R}^{m\times m}$
and $\bm{C}_{\eta}(\tau)\in\mathbb{R}^{m\times m}$ as
$\bm{C}_{\xi}(\tau)=\langle \bm{\xi}(t)\bm{\xi}(t-\tau)^{\top}\rangle$
and $\bm{C}_{\eta}(\tau)=\langle \bm{\eta}^{(j)}(t)\bm{\eta}^{(j)}(t-\tau)^{\top}
\rangle$ ($j=1,\ldots,N$)}.
{For the sake of simplicity, we assumed that
all independent noises $\bm{\eta}^{(j)}(t)$ have
the same statistical property
characterized by $\bm{C}_{\eta}(\tau)$.}
The ($i$, $j$)th element of $\bm{C}_{\xi}(\tau)$
is the cross correlation function
of the $i$th and $j$th elements of the common noise $\bm{\xi}(t)$.
The diagonal elements of $\bm{C}_{\xi}(\tau)$
are autocorrelation functions.
{In the same way, we can characterize
the statistical property of $\bm{\eta}^{(j)}(t)$
by using $\bm{C}_{\eta}(\tau)$}.

\section{Phase reduction}

Under the assumption that the noise intensity
is sufficiently weak ($D\ll1$ and $\epsilon\ll1$),
we can apply the phase reduction method
\cite{kuramoto03__chemic_oscil_waves_and_turbul,
goldobin10__dynam_of_limit_cycle_oscil}
to eq. (\ref{eq. lim osci}).
By introducing a phase variable $\phi^{(j)}$,
eq. (\ref{eq. lim osci}) is reduced
to the following phase equation:
\begin{eqnarray}
 \dot{\phi}^{(j)} &=& \omega +
	\sqrt{D}\bm{Z}_{\rm G}(\phi^{(j)})\cdot\bm{\xi}(t)
	\nonumber \\
	&&+\sqrt{\epsilon}\bm{Z}_{\rm H}(\phi^{(j)})\cdot\bm{\eta}^{(j)}(t)
	 + O(D,\epsilon),
	\label{eq. ph osci}
\end{eqnarray}
where $\phi^{(j)}(t)\in[-\pi,+\pi]$ is a phase variable
that corresponds to the state of the $j$th oscillator
$\bm{X}^{(j)}$,
$\omega$ ($=2\pi T^{-1}$) is the natural frequency,
and $\bm{Z}_G(\phi^{(j)})$ and $\bm{Z}_H(\phi^{(j)})$
are the phase sensitivity functions
that represent the linear response of the phase variable $\phi^{(j)}$
to the drive noises\cite{kuramoto03__chemic_oscil_waves_and_turbul,
goldobin10__dynam_of_limit_cycle_oscil}.
{The phase sensitivity functions
$\bm{Z}_G(\phi^{(j)})$ and $\bm{Z}_H(\phi^{(j)})$
are defined as $\bm{Z}_G(\phi^{(j)})
=\nabla_{\bm{X}}\phi^{(j)}|_{\bm{X}=\bm{S}(\phi^{(j)})}
\cdot\bm{G}(\bm{S}(\phi^{(j)}))$
and $\bm{Z}_H(\phi^{(j)})
=\nabla_{\bm{X}}\phi^{(j)}|_{\bm{X}=\bm{S}(\phi^{(j)})}
\cdot\bm{H}(\bm{S}(\phi^{(j)}))$}.
As discussed in Ref.
\cite{goldobin10__dynam_of_limit_cycle_oscil},
the $O(D,\epsilon)$ term is necessary
to describe the exact phase dynamics,
while the phase diffusion is not affected
by the $O(D,\epsilon)$ term.
As we will focus on the phase diffusion
in the following sections,
we do not take this term into account.

\section{Effective Langevin description}

To quantify the synchronization property
without loss of generality,
we consider {the relationship of only two oscillators,
that is,} the two-body problem
of ${\phi}^{(1)}(t)$ and ${\phi}^{(2)}(t)$,
and define the phase difference $\theta$
($:={\phi}^{(1)}-{\phi}^{(2)}$).
As we focus on the {stochastic} dynamics of $\theta$,
we define $f(\theta,t)$ as the PDF of the phase difference $\theta$.
Utilizing the effective white-noise Langevin description
\cite{nakao10__effec_long_time_phase_dynam},
the evolution of $f(\theta,t)$ is described by the
following effective Fokker-Planck equation:
\begin{eqnarray}
 \frac{\partial f}{\partial t}
	+\frac{\partial}{\partial\theta}v^{(1)}(\theta)f
	-\frac{1}{2}\cdot\frac{\partial^2}{\partial\theta^2}v^{(2)}(\theta)f=0,
	\label{eq. eff fp}
\end{eqnarray}
where $v^{(1)}(\theta)$ and $v^{(2)}(\theta)$ are effective
drift and diffusion coefficients.
We have the drift coefficient $v^{(1)}(\theta)=0$ because
$\langle \dot{\theta} \rangle = \langle \dot{{\phi}}^{(1)}-
\dot{{\phi}}^{(2)} \rangle = 0$.
Meanwhile, the diffusion coefficient $v^{(2)}(\theta)$ is obtained as
\begin{eqnarray}
 v^{(2)}(\theta) &=& \int_{-\infty}^{+\infty}d\tau
	\big\langle \big[\dot{\theta}(t)-\langle\dot{\theta}\rangle\big]
	\big[\dot{\theta}(t-\tau)-\langle\dot{\theta}\rangle\big] \big\rangle
	\nonumber \\
	&=& \int_{-\infty}^{+\infty}d\tau
	\big\langle \big[\dot{{\phi}}^{(1)}(t)-\dot{{\phi}}^{(2)}(t)\big]
	\nonumber \\
	&& \ \ \big[\dot{{\phi}}^{(1)}(t-\tau)-\dot{{\phi}}^{(2)}(t-\tau)\big]
	\big\rangle
	\label{eq. diff coeff2}
\end{eqnarray}
where $\langle\cdot\rangle$ represents the temporal average.
{For simplicity of notation, we define $d_{jk}$ as
{$d_{jk} = \int_{-\infty}^{+\infty}d\tau
\big\langle[\dot{\phi}^{(j)}(t)-\omega]
[\dot{\phi}^{(k)}(t-\tau)-\omega]\big\rangle$}.
Then, we obtain
\begin{eqnarray}
 v^{(2)}(\theta) = d_{11} + d_{22} - d_{12} - d_{21}
	= 2d_{11}-2d_{12}.
	\label{eq. diff coeff}
\end{eqnarray}}

The phase variable $\phi^{(j)}(t)$ can be
expanded as $\phi^{(j)}(t)=\phi^{(j)}_0(t)
+\sqrt{D}\phi^{(j)}_{D,1}(t)+\sqrt{\epsilon}\phi^{(j)}_{\epsilon,1}(t)
+D\phi^{(j)}_{D,2}(t)+\epsilon\phi^{(j)}_{\epsilon,2}(t)
+\cdots$ by using $\sqrt{D}$
and $\sqrt{\epsilon}$ as expansion parameters,
where $\phi^{(j)}_0(t)$, $\phi^{(j)}_{D,k}(t)$
and $\phi^{(j)}_{\epsilon,k}(t)$
($k=1,2,\ldots$) are approximate perturbed solutions of $\phi^{(j)}(t)$.
We have $\phi^{(j)}_0(t)=\phi^{(j)}_0(0)+\omega t$,
$\dot{\phi}^{(j)}_{D,1}(t)=\bm{Z}_G(\phi^{(j)}_0(t))\cdot\bm{\xi}(t)$
and $\dot{\phi}^{(j)}_{\epsilon,1}(t)=\bm{Z}_H(\phi^{(j)}_0(t))\cdot\bm{\eta}^{(j)}(t)$.
{
Using these perturbed solutions,
eq. (\ref{eq. ph osci}) can be written
as $\dot{\phi}^{(j)}=\omega+\sqrt{D}\dot{\phi}^{(j)}_{D,1}
+\sqrt{\epsilon}\dot{\phi}^{(j)}_{\epsilon,1}+O(D,\epsilon)$.}
Using this approximation and the fact that
$\langle\phi^{(j)}_{D,1}(t)\phi^{(k)}_{\epsilon,1}(t-\tau)\rangle=0$
and $\langle\phi^{(j)}_{\epsilon,1}(t)\phi^{(k)}_{D,1}(t-\tau)\rangle=0$,
we obtain
\begin{eqnarray}
 d_{jk}&=&D\int_{-\infty}^{+\infty}d\tau
	\big\langle\dot{\phi}^{(j)}_{D,1}(t)\dot{\phi}^{(k)}_{D,1}(t-\tau)\big\rangle
	\nonumber \\
	&& + \epsilon\int_{-\infty}^{+\infty}d\tau
	\big\langle\dot{\phi}^{(j)}_{\epsilon,1}(t)\dot{\phi}^{(k)}_{\epsilon,1}(t-\tau)\big\rangle
	\nonumber \\
	&& + O(D^{\frac{3}{2}},\epsilon^{\frac{3}{2}}).
	 \label{eq. d app}
\end{eqnarray}

Thus, using eq. (\ref{eq. d app}),
we can calculate $d_{11}$ as follows:
\begin{eqnarray}
 d_{11}
 &=& \frac{D}{2\pi} \int_{-\infty}^{+\infty}d\tau
	\int_{-\pi}^{+\pi}d\phi
	\nonumber \\
 && \ \ \bm{Z}_G(\phi)^{\top}\bm{C}_{\xi}(\tau)\bm{Z}_G(\phi-\omega\tau)
	\nonumber \\
 && + \frac{\epsilon}{2\pi} \int_{-\infty}^{+\infty}d\tau\int_{-\pi}^{+\pi}d\phi
	\nonumber \\
 && \ \ \bm{Z}_H(\phi)^{\top}\bm{C}_{\eta}(\tau)\bm{Z}_H(\phi-\omega\tau)
	+ O(D^{\frac{3}{2}},\epsilon^{\frac{3}{2}}).
	\label{eq. calc d11}
\end{eqnarray}
In the same way, $d_{12}$ is given by
\begin{eqnarray}
 d_{12}
 &=& \frac{D}{2\pi} \int_{-\infty}^{+\infty}d\tau
	\int_{-\pi}^{+\pi}d\phi
	\nonumber \\
 && \ \
	\bm{Z}_G(\phi)^{\top}\bm{C}_{\xi}(\tau)\bm{Z}_G(\phi-\theta-\omega\tau)
	\nonumber \\
 && + O(D^{\frac{3}{2}},\epsilon^{\frac{3}{2}}).
	\label{eq. calc d12}
\end{eqnarray}
The detailed derivations of eqs. (\ref{eq. calc d11}) and (\ref{eq. calc d12})
are shown in Appendix A.

Finally, from eqs. (\ref{eq. diff coeff}), (\ref{eq. calc d11})
and (\ref{eq. calc d12}),
we have the efficient diffusion
coefficient $v^{(2)}(\theta)$:
\begin{eqnarray}
 v^{(2)}(\theta) = 2D\big[g(0) - g(\theta)\big]
	+ 2\epsilon h(0),
	\label{eq. v2}
\end{eqnarray}
where $g(\theta)$ and $h(\theta)$
are correlation functions defined as
\begin{eqnarray}
 g(\theta) &=& \frac{1}{2\pi} 
	\int_{-\infty}^{+\infty} d\tau \int_{-\pi}^{+\pi} d\phi \nonumber \\
 && \ \ \bm{Z}_G(\phi)^{\top} \bm{C}_{\xi}(\tau)
	\bm{Z}_G(\phi - \theta - \omega \tau), \label{eq. corr g} \\
 h(\theta) &=& \frac{1}{2\pi} 
	\int_{-\infty}^{+\infty} d\tau \int_{-\pi}^{+\pi} d\phi \nonumber \\
 && \ \ \bm{Z}_H(\phi)^{\top} \bm{C}_{\eta}(\tau)
	\bm{Z}_H(\phi - \theta - \omega \tau).
	\label{eq. corr h}
\end{eqnarray}
If we assume that the drive noise is white, namely,
$\bm{C}_{\xi}(\tau)=\bm{C}_{\eta}(\tau)=\delta(\tau)\bm{E}_m$,
eqs. (\ref{eq. corr g}) and (\ref{eq. corr h}) are exactly equivalent to
eq. (6) in Ref. \cite{nakao07__noise_induc_synch_and_clust},
where $\bm{E}_m$ is an $m\times m$ identity matrix.
The results show that eqs. (\ref{eq. corr g}) and (\ref{eq. corr h}) are a natural generalization
of eq. (6) in Ref. \cite{nakao07__noise_induc_synch_and_clust}.

We obtain the explicit form of the Fokker-Planck equation
of eq. (\ref{eq. eff fp})
from eqs. (\ref{eq. v2})--(\ref{eq. corr h}).
The stationary distribution of the phase difference
$f_0(\theta)$ is given as
the stationary solution of eq. (\ref{eq. eff fp}).
Then, if we put $\partial f/\partial t=0$ in eq. (\ref{eq. eff fp}),
we obtain
{
\begin{eqnarray}
 f_0(\theta) = \frac{\nu}{v^{(2)}(\theta)}
	=\frac{\nu'}{D\big[g(0)-g(\theta)\big]+\epsilon h(0)},
\end{eqnarray}
}
where $\nu$ and $\nu'$ ($=\nu/2$) are normalization constants.

\section{Fourier representation}

To understand the results obtained
in the previous section,
we rewrite the correlation functions
defined in eqs. (\ref{eq. corr g}) and (\ref{eq. corr h})
by using the Fourier representation.
We introduced the Fourier series expansion
of the phase sensitivity functions
$\bm{Z}_G(\phi)$ and $\bm{Z}_H(\phi)$ as
{
$\bm{Z}_G(\phi) = \sum_{l=-\infty}^{+\infty}\bm{Y}_{G,l}e^{il\phi}$ and 
$\bm{Z}_H(\phi) = \sum_{l=-\infty}^{+\infty}\bm{Y}_{H,l}e^{il\phi}$
},
where $i$ denotes the imaginary unit
and $\bm{Y}_{G,l}\in\mathbb{C}^m$
($=\frac{1}{2\pi}\int_{-\pi}^{+\pi}d\phi
\bm{Z}_G(\phi)e^{-il\phi}$)
and $\bm{Y}_{H,l}\in\mathbb{C}^m$
($=\frac{1}{2\pi}\int_{-\pi}^{+\pi}d\phi
\bm{Z}_H(\phi)e^{-il\phi}$)
are Fourier coefficients
($l=-\infty,\ldots,\infty$).

Subsequently, we define $\bm{P}_{\xi}(\Omega)\in\mathbb{C}^{m\times m}$
and $\bm{P}_{\eta}(\Omega)\in\mathbb{C}^{m\times m}$
as the Fourier transforms of
$\bm{C}_{\xi}(\tau)$ and $\bm{C}_{\eta}(\tau)$,
{that is,
$\bm{P}_{\xi}(\Omega) = \int_{-\infty}^{+\infty} dt
	\bm{C}_{\xi}(t)e^{-i\Omega t}$ and
$\bm{P}_{\eta}(\Omega) = \int_{-\infty}^{+\infty} dt
	\bm{C}_{\eta}(t)e^{-i\Omega t}$}.
Let us note that $\bm{P}_{\xi}(\Omega)$ and $\bm{P}_{\eta}(\Omega)$
are Hermitian matrices, namely,
$\bm{P}_{\xi}(\Omega)=\bm{P}_{\xi}(\Omega)^{\dag}$ and
$\bm{P}_{\eta}(\Omega)=\bm{P}_{\eta}(\Omega)^{\dag}$
because $\bm{C}_{\xi}(\tau)=\bm{C}_{\xi}(-\tau)^{\top}$
and $\bm{C}_{\eta}(\tau)=\bm{C}_{\eta}(-\tau)^{\top}$
from their definitions, where
$\dag$ denotes the adjoint.
The $(i,j)$th elements of
$\bm{P}_{\xi}(\Omega)$ and $\bm{P}_{\eta}(\Omega)$
represent the cross spectra of
the $i$th and $j$th elements of $\bm{\xi}(t)$ and
$\bm{\eta}^{(j)}(t)$.
In particular, the diagonal elements of $\bm{P}_{\xi}(\Omega)$ and
$\bm{P}_{\eta}(\Omega)$ represent the power spectra.

Using the Fourier representations
defined {above},
we can obtain the Fourier representations
of the correlation functions $g(\theta)$ and $h(\theta)$:
\begin{eqnarray}
 g(\theta) = \sum_{l=-\infty}^{+\infty}g_le^{il\theta},\ 
 h(\theta) = \sum_{l=-\infty}^{+\infty}h_le^{il\theta},
	\label{eq. corr fr}
\end{eqnarray}
where $g_l$ ($=\bm{Y}_{G,l}^{\dag}\bm{P}_{\xi}(l\omega)\bm{Y}_{G,l}$)
and $h_l$ ($=\bm{Y}_{H,l}^{\dag}\bm{P}_{\eta}(l\omega)\bm{Y}_{H,l}$)
are Fourier coefficients
($l=-\infty,\ldots,\infty$).
The derivations of $g_l$ and $h_l$ will be shown
in Appendix B.

These expressions clearly suggest that
the correlation functions $g(\theta)$ and $h(\theta)$
only depend on $\bm{P}_{\xi}(\pm l\omega)$
and $\bm{P}_{\eta}(\pm l\omega)$
($l=0,1,2,\ldots$), that is,
the other frequency components can be neglected.
In the next section, we will demonstrate
that colored noise induces various synchronized and clustered states,
which are clearly explained by eq. (\ref{eq. corr fr}).

\section{Numerical simulations}

\begin{figure}[tbp]
 \begin{center}
  \includegraphics[bb=0 0 600 570,width=80mm,clip]{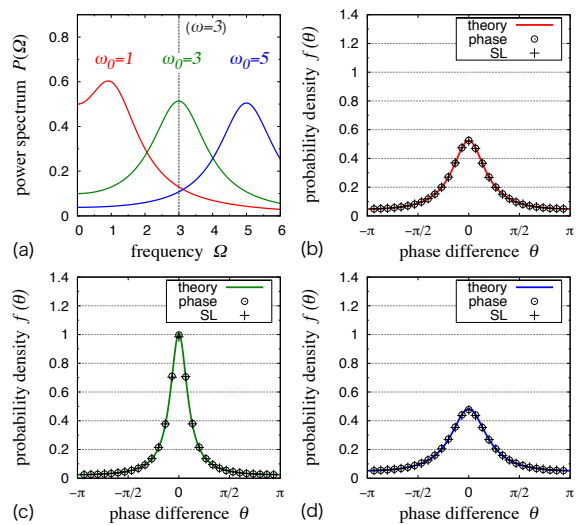}
 \end{center}
 \caption{(Color online) Simulation results of the Stuart-Landau oscillator (crosses)
 and the corresponding phase oscillator (open circles).
 (a) Power spectra of the common noises are
 shown for $\omega_0=1$, 3 and 5.
 The PDFs of $\theta$ show the frequency dependency of the
 synchronization property for
 (b) $\omega_0=1$, (c) $\omega_0=3$ ($=\omega$), and (d) $\omega_0=5$.}
 \label{fig:one}
\end{figure}

To demonstrate the validity of our results,
we perform numerical experiments for two types of
limit cycle oscillators.
The first example is the Stuart-Landau oscillator,
which takes the normal form of
the supercritical Hopf bifurcation
\cite{kuramoto03__chemic_oscil_waves_and_turbul}:
{
$\dot{x} = x - c_0y - (x^2 + y^2)(x - c_2y)$,
$\dot{y} = y + c_0x - (x^2 + y^2)(y + c_2x)$},
where $\bm{X}=[x,y]^{\top}$
is a state variable
and $c_0$ and $c_2$ are parameters.
In the simulation, we fixed
$c_0=1$, $c_2=-2$, $\bm{G}=\bm{H}={\rm diag}(1,1)$,
$D=0.0095$ and $\epsilon=0.0005$,
where ${\rm diag}(\lambda_1,\ldots,\lambda_m)$ denotes
an $m\times m$ diagonal matrix
that has the diagonal elements $\lambda_1,\ldots,\lambda_m$.
{This model is}
reduced to the phase equation that has
the natural frequency $\omega=c_0-c_2=3$ and
the phase sensitivity function
$\bm{Z}(\phi)=\sqrt{2}[\sin(\phi+3\pi/4),\sin(\phi+\pi/4)]^{\top}$.

In the simulation, we use a two-dimensional drive noise that
has the correlation matrix $\bm{C}_{\rm ex}(\tau)\in\mathbb{R}^{2\times2}$
defined as{
$\bm{C}_{\rm ex}(\tau)={\rm diag}(C_{\rm ex}(\tau),C_{\rm ex}(\tau))$
and $C_{\rm ex}(\tau)=\frac{\gamma}{2}e^{-\gamma|\tau|}\cos\omega_0\tau$},
where $\omega_0$ and $\gamma$ are parameters
that represent the peak frequency and the characteristic decay time.
We define $P_{\rm ex}(\Omega)$, the Fourier transform of
$C_{\rm ex}(\tau)$, as{
$P_{\rm ex}(\Omega)
=\frac{\gamma^2}{2}\{[\gamma^2+(\Omega+\omega_0)^2]^{-1}
+[\gamma^2+(\Omega-\omega_0)^2]^{-1}\}$}.
{A drive noise characterized by $C_{\rm ex}(\tau)$}
can be generated by the damped noisy harmonic oscillator
(See eqs. (43)--(49) in Ref. \cite{nakao10__effec_long_time_phase_dynam}
for details).

We use the common noises with
$(\omega_0,\gamma)=(1,1)$, $(3,1)$ and $(5,1)$
and the independent noise with
$(\omega_0,\gamma)=(0,3)$.
The power spectra of these common noises
are shown in fig. \ref{fig:one} (a).
From eq. (\ref{eq. corr fr}),
the correlation functions $g(\theta)$ and $h(\theta)$
are given by{
$g(\theta)=\{[1+(\omega_0+3)^2]^{-1}
+[1+(\omega_0-3)^2]^{-1}\}\cos\theta$
and $h(\theta)= \cos\theta$},
for $\omega_0 =$ 1, 3 and 5, which correspond to
the three types of the common noise.
The derivations of $g(\theta)$ and $h(\theta)$
will be shown in Appendix C.

{The correlation function $g(\theta)$ calculated above}
indicate that the effective intensity of the common noise
depends on the peak frequency $\omega_0$
and is maximal at $\omega_0=\omega$.
It means that the synchronous degree is maximized at $\omega_0=\omega$.
In fig. \ref{fig:one} (b)--(d), we compared the results of
the direct numerical simulation
using the Stuart-Landau oscillator and its corresponding phase
oscillator with the analytical results.
All PDFs are well fitted by the theoretical curves.
Our theory clearly predicts that the highest synchronous degree is realized
at $\omega_0=\omega$.

\begin{figure}[tbp]
 \begin{center}
  \includegraphics[bb=0 0 600 570,width=80mm,clip]{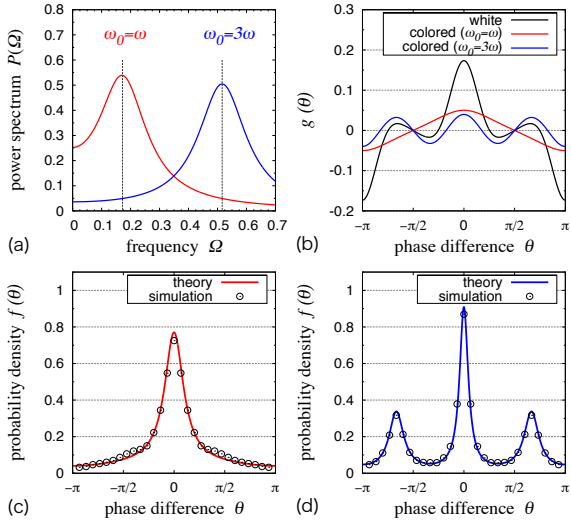}
 \end{center}
 \caption{(Color online) Simulation results of the
 FitzHugh-Nagumo oscillator.
 (a) Power spectra of the common noises
 are shown for $\omega_0=\omega$ and $3\omega$.
 For these drive noises, (b) $g(\theta)$ ($=h(\theta)$) is shown.
 The PDFs of $\theta$ for
 (c) $\omega_0=\omega$ (synchronized state)
 and for (d) $\omega_0=3\omega$ (3-cluster state)
 are shown.}
 \label{fig:two}
\end{figure}

The second example is the FitzHugh-Nagumo
oscillator
\cite{fhn,nagumo}:
{
$\dot{v}=v-v^3/3-u+I_0$,
$\dot{u}=\mu(v+a-bu)$},
where $\bm{X}=[v,u]^{\top}$ is a state variable
and $a$, $b$, $\mu$ and $I_0$ are parameters.
In the simulation,
we fixed $a=0.7$, $b=0.8$, $\mu=0.08$, $I_0=0.875$,
$\bm{G}=\bm{H}={\rm diag}(1,0)$,
$D=0.045$ and $\epsilon=0.005$.
For these parameters, this oscillator has
the natural frequency {$\omega\simeq0.1725$}.
This oscillator models bursting behavior of a neuron,
and only the first variable $v$,
which corresponds to the membrane potential of a neuron,
is subject to noise.

In the simulation, we use the one-dimensional
noise that has the correlation function
$C_{\rm ex}(\tau)$.
Different from the first example, we use
the same parameters $(\omega_0,\gamma)$
for both the common and independent noises.
We used two parameter sets $(\omega_0,\gamma)=(\omega,0.1)$ and $(3\omega,0.1)$.
The power spectra of these drive noises
are shown in fig. \ref{fig:two} (a).
We obtain the correlation function $g(\theta)$ ($=h(\theta)$)
numerically as shown in fig. \ref{fig:two} (b).

In fig. \ref{fig:two} (c) and (d),
we compared the results of
the direct numerical simulation
with the analytical results.
The numerical results are in good agreement
with the theoretical results.
As theoretically predicted, a 3-cluster state is realized
as shown in fig. \ref{fig:two} (d).
{If oscillators are driven by white noise,
clustered states are induced only by multiplicative noise
\cite{nakao07__noise_induc_synch_and_clust}.
However, in case of colored noise,
clustered states are induced not only by multiplicative noise
but also by additive noise}.

\begin{figure}[tbp]
 \begin{center}
  \includegraphics[width=80mm,clip]{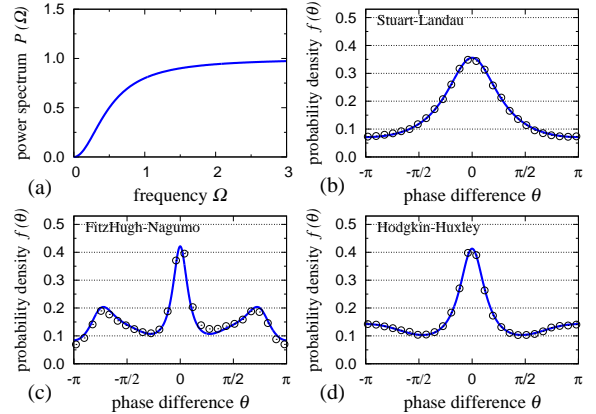}
 \end{center}
 \caption{{(Color online) Simulation results of
 the limit cycle oscillators subject to {green noise}.
 (a) Power spectra of the drive noises.
 The PDFs of $\theta$ obtained by the theory (lines)
 and numerical simulations (circles)
 for (b) the Stuart-Landau oscillator,
 (c) the FitzHugh-Nagumo oscillator and
 (d) the Hodgkin-Huxley oscillator.}}
 \label{fig:three}
\end{figure}

{
In the third example,
we used the Hodgkin-Huxley oscillator\cite{hh},
which enables us to demonstrate whether the theory is applicable to
higher-dimensional limit cycle systems.
We use
{green noise
used in ref. \cite{wang},
which is} generated by applying a high-pass filter to white noise.
{The power spectrum is} shown in fig. \ref{fig:three} (a).
Different from the periodic noise characterized by $C_{\rm ex}(\tau)$,
the {green noise} has a vanishing spectrum
as $\Omega\to0$.
In the simulation, for the sake of simplicity,
we used the same type of drive noise
for the common and independent noises,
and we set the noise intensities $(D,\epsilon)=(0.0002,0.0001)$.
Fig. \ref{fig:three} (b)--(d) compare
the theoretical and numerical results,
which show that
our theory is also valid for these cases.
}

\section{Summary and discussions}

In this Letter, we extended a formulation
to analyze various synchronized and clustered states
of uncoupled limit cycle oscillators driven by
common and independent colored noises.
Using this formulation, we derived the
probability density function of the
phase difference and rewrote it
by the Fourier representation.
The obtained expressions
clearly show that
the synchronization property depends
on the power spectrum of the drive noises.
Such dependency has already been reported experimentally.
For example,
in ref. \cite{fellous01__frequen_depen_of_spike_timin},
the reliability, or synchronization across trials,
is explored in neuronal responses to periodic drive inputs
with various frequencies.
The reliability is maximized at a certain frequency,
{which is similar to
our results shown in fig. \ref{fig:one}}.
Our results in this Letter supports the results in
ref. \cite{fellous01__frequen_depen_of_spike_timin}
theoretically, because
a neuron in a oscillatory state can be regarded as
a noisy limit cycle oscillator.

Generally, noise in the real world often has
a non-flat and characteristic power spectrum.
In this sense,
our formulation is a useful tool to estimate
the synchronization property
for both theoretical and practical aspects.
Namely, the results obtained in this Letter
can be applied to a wide range of purposes
from mathematical modelings
to technological problems.

\acknowledgments
The authors would like to thank S. Ogawa and AGS
Corp.~for their encouragement on this research project.

\section{Appendix A: Derivations of eqs. (\ref{eq. calc d11}) and (\ref{eq. calc d12})}
\renewcommand{\theequation}{A.\arabic{equation} }
\setcounter{equation}{0}

Substituting
$\dot{\phi}_{D,1}^{(j)}=
\bm{Z}_G(\phi_0^{(j)}(t))\cdot\bm{\xi}(t)$
and 
$\dot{\phi}_{\epsilon,1}^{(j)}=
\bm{Z}_H(\phi_0^{(1)}(t))\cdot\bm{\eta}^{(1)}(t)$
into eq. (\ref{eq. d app}), we obtain
\begin{eqnarray}
 d_{11} &=& D
	\int_{-\infty}^{+\infty}d\tau
	\big\langle [\bm{Z}_G(\phi_0^{(1)}(t))^{\top}\bm{\xi}(t)]
	\nonumber \\
 &&\ \ [\bm{Z}_G(\phi_0^{(1)}(t-\tau))^{\top}\bm{\xi}(t-\tau)]\big\rangle
  \nonumber \\
 && + \epsilon \int_{-\infty}^{+\infty}d\tau
	\big\langle [\bm{Z}_H(\phi_0^{(1)}(t))^{\top}\bm{\eta}^{(1)}(t)]
	\nonumber \\
 &&\ \ [\bm{Z}_H(\phi_0^{(1)}(t-\tau))^{\top}\bm{\eta}^{(1)}(t-\tau)]\big\rangle
	\nonumber \\	
 && + O(D^{\frac{3}{2}},\epsilon^{\frac{3}{2}}) \\
 &=& D
	\int_{-\infty}^{+\infty}d\tau
	\big\langle \bm{Z}_G(\phi_0^{(1)}(t))^{\top}\bm{\xi}(t)
	\nonumber \\
 && \ \ \bm{\xi}(t-\tau)^{\top}\bm{Z}_G(\phi_0^{(1)}(t-\tau)) \big\rangle
	\nonumber \\ 
 && + \epsilon
	\int_{-\infty}^{+\infty}d\tau
	\big\langle \bm{Z}_H(\phi_0^{(1)}(t))^{\top}\bm{\eta}^{(1)}(t)
	\nonumber \\
 && \ \ \bm{\eta}^{(1)}(t-\tau)^{\top}\bm{Z}_H(\phi_0^{(1)}(t-\tau))\big\rangle
	\nonumber \\
 && + O(D^{\frac{3}{2}},\epsilon^{\frac{3}{2}}).
\end{eqnarray}
We rewrite $\bm{Z}_G(\phi)$, $\bm{Z}_H(\phi)$, $\bm{\xi}(t)$
and $\bm{\eta}^{(1)}(t)$ by using their elements and obtain

\begin{eqnarray}
 d_{11}
 &=& D \int_{-\infty}^{+\infty}d\tau
	\sum_{k=1}^m\sum_{l=1}^m \big\langle Z_{H,k}(\phi_0^{(1)}(t))
	\nonumber \\
 && \ \ \xi_k(t)
	\xi_l(t-\tau)Z_{H,l}(\phi_0^{(1)}(t-\tau))
	\big\rangle
	\nonumber \\
 && + \epsilon \int_{-\infty}^{+\infty}d\tau\
	\sum_{k=1}^m\sum_{l=1}^m \big\langle Z_{H,k}(\phi_0^{(1)}(t))
	\nonumber \\
 && \ \ \eta^{(1)}_k(t)
	\eta^{(1)}_l(t-\tau)Z_{H,l}(\phi_0^{(1)}(t-\tau))
	\big\rangle
 \nonumber \\
 && + O(D^{\frac{3}{2}},\epsilon^{\frac{3}{2}}),
\end{eqnarray}
where $Z_{G,l}(\phi)$ and $Z_{H,l}(\phi)$ are the $l$th elements of
$\bm{Z}_G(\phi)$ and $\bm{Z}_H(\phi)$, and
$\xi_l(t)$ and $\eta_l^{(1)}(t)$ are the $l$th elements of
$\bm{\xi}(t)$ and $\bm{\eta}^{(1)}(t)$.

We assume that the phase variable $\phi^{(1)}$
and the drive noises $\bm{\xi}(t)$ and $\bm{\eta}^{(1)}(t)$
are approximately independent.
Under this assumption,
the temporal average $\langle\cdot\rangle$
can be divided into two parts;
$\langle\cdot\rangle_{\phi}$
($:=(2\pi)^{-1}\int_{-\pi}^{+\pi}d\phi\,\cdot\,$)
and $\langle\cdot\rangle_t$
({$:=\lim_{s \to \infty}(2s)^{-1}\int_{-s}^{+s} dt\,\cdot\,$}).
Thus, we obtain
\begin{eqnarray}
 d_{11}
 &=& D \int_{-\infty}^{+\infty}d\tau
	\sum_{k=1}^m\sum_{l=1}^m \big\langle Z_{G,k}(\phi_0^{(1)}(t))
	\nonumber \\
 && \ \ Z_{G,l}(\phi_0^{(1)}(t-\tau))
	\big\rangle_{\phi}
	\big\langle\xi_k(t)\xi_l(t-\tau)
	\big\rangle_t
	\nonumber \\
 && + \epsilon\int_{-\infty}^{+\infty}d\tau\
	\sum_{k=1}^m\sum_{l=1}^m \big\langle Z_{H,k}(\phi_0^{(1)}(t))
	\nonumber \\
 && \ \ Z_{H,l}(\phi_0^{(1)}(t-\tau))
	\big\rangle_{\phi}
	\big\langle\eta^{(1)}_k(t)\eta^{(1)}_l(t-\tau)
	\big\rangle_t
 \nonumber \\
 && + O(D^{\frac{3}{2}},\epsilon^{\frac{3}{2}})
	\nonumber \\
	&=& \frac{D}{2\pi} \int_{-\infty}^{+\infty}d\tau
	\int_{-\pi}^{+\pi}d\phi
	\nonumber \\
 && \ \ \sum_{k=1}^m\sum_{l=1}^m Z_{G,k}(\phi)Z_{G,l}(\phi-\omega\tau)
	C_{\xi,kl}(\tau)
	\nonumber \\
 && + \frac{\epsilon}{2\pi}\int_{-\infty}^{+\infty}d\tau\
 \int_{-\pi}^{+\pi}d\phi
	\nonumber \\
 && \ \ \sum_{k=1}^m\sum_{l=1}^m Z_{H,k}(\phi)Z_{H,l}(\phi-\omega\tau)
	C_{\eta,kl}(\tau)
 \nonumber \\
 && + O(D^{\frac{3}{2}},\epsilon^{\frac{3}{2}}),
	\label{eq. app elem}
\end{eqnarray}
where $C_{\xi,kl}$ and $C_{\eta,kl}$ are the $(k,l)$th elements of
$\bm{C}_{\xi}(\phi)$ and $\bm{C}_{\eta}(\phi)$.
Finally, we rewrite eq. (\ref{eq. app elem})
by using $\bm{Z}_G(\phi)$, $\bm{Z}_H(\phi)$,
$\bm{C}_{\xi}(\tau)$ and $\bm{C}_{\eta}(\tau)$ and obtain
\begin{eqnarray}
 d_{11}
 &=& \frac{D}{2\pi} \int_{-\infty}^{+\infty}d\tau
	\int_{-\pi}^{+\pi}d\phi
	\nonumber \\
 && \ \ \bm{Z}_G(\phi)^{\top}\bm{C}_{\xi}(\tau)\bm{Z}_G(\phi-\omega\tau)
	\nonumber \\
 && + \frac{\epsilon}{2\pi} \int_{-\infty}^{+\infty}d\tau\int_{-\pi}^{+\pi}d\phi
	\nonumber \\
 && \ \ \bm{Z}_H(\phi)^{\top}\bm{C}_{\eta}(\tau)\bm{Z}_H(\phi-\omega\tau)
	\nonumber \\
 && + O(D^{\frac{3}{2}},\epsilon^{\frac{3}{2}}).
\end{eqnarray}

In the same way, one can calculate $d_{12}$ as follows.
We use the fact that
$\langle\phi^{(1)}_{\epsilon,1}(t)\phi^{(2)}_{\epsilon,1}(t-\tau)\rangle=0$
and eliminate the phase variable of the second oscillator $\phi^{(2)}_0$
by substituting $\phi^{(2)}_0=\phi^{(1)}_0-\theta$ into $\phi^{(2)}_0$,
and then, we obtain
\begin{eqnarray}
 d_{12} &=& D
	\int_{-\infty}^{+\infty}d\tau
	\big\langle [\bm{Z}_G(\phi_0^{(1)}(t))^{\top}\bm{\xi}(t)]
	\nonumber \\
 &&\ \ [\bm{Z}_G(\phi_0^{(2)}(t-\tau))^{\top}\bm{\xi}(t-\tau)]\big\rangle
  \nonumber \\
 && + O(D^{\frac{3}{2}},\epsilon^{\frac{3}{2}}) \nonumber \\
 &=& D
	\int_{-\infty}^{+\infty}d\tau
	\big\langle \bm{Z}_G(\phi_0^{(1)}(t))^{\top}\bm{\xi}(t)
	\nonumber \\ 
 &&\ \ \bm{\xi}(t-\tau)^{\top}\bm{Z}_G(\phi_0^{(2)}(t-\tau))\big\rangle
  \nonumber \\
 && + O(D^{\frac{3}{2}},\epsilon^{\frac{3}{2}}) \nonumber \\
 &=& \frac{D}{2\pi} \int_{-\infty}^{+\infty}d\tau
	\int_{-\pi}^{+\pi}d\phi
	\nonumber \\
 && \ \
	\bm{Z}_G(\phi)^{\top}\bm{C}_{\xi}(\tau)\bm{Z}_G(\phi-\theta-\omega\tau)
	\nonumber \\
 && + O(D^{\frac{3}{2}},\epsilon^{\frac{3}{2}}).
\end{eqnarray}

\section{Appendix B: Derivation of eq. (\ref{eq. corr fr})}
\renewcommand{\theequation}{B.\arabic{equation} }
\setcounter{equation}{0}

From eq. (\ref{eq. corr g}), one can calculate the Fourier coefficient
$g_l$ as follows.
We introduce a new variable $\chi$
($:=\phi-\theta-\omega\tau$)
and use the fact that $\bm{P}_{\xi}(\Omega)$ is a Hermitian
matrix. Then, we obtain
\begin{eqnarray}
 g_l &=& \frac{1}{2\pi}\int_{-\pi}^{+\pi}d\theta g
	(\theta)e^{-i l\theta}
	\nonumber \\
 &=& \frac{1}{2\pi}\int_{-\pi}^{+\pi}d\theta \frac{1}{2\pi}
	\int_{-\infty}^{+\infty}d\tau
	\int_{-\pi}^{+\pi}d\phi \bm{Z}_G(\phi)^{\top}
	\nonumber \\
 &&\ \ \bm{C}_{\xi}(\tau)\bm{Z}_G(\phi-\theta-\omega\tau) e^{-i l\theta}
	\nonumber \\
 &=& \bigg(\frac{1}{2\pi}\int_{-\pi}^{+\pi}d\phi
	\bm{Z}_G(\phi)^{\top} e^{-il\phi} \bigg)
	\nonumber \\
 && \bigg( \int_{-\infty}^{+\infty}d\tau
	\bm{C}_{\xi}(\tau) e^{il \omega\tau} \bigg)
	\bigg(\frac{1}{2\pi}\int_{-\pi}^{+\pi}d\chi
	 \bm{Z}_G(\chi)e^{il\chi} \bigg)
	\nonumber \\
	&=& \bm{Y}_{G,l}^{\top}
	 \overline{\bm{P}_{\xi}(l\omega)}\,\overline{\bm{Y}_{G,l}}
	= \bm{Y}_{G,l}^{\dag}
	\bm{P}_{\xi}(l\omega)^{\dag} \bm{Y}_{G,l}
	\nonumber \\
 &=& \bm{Y}_{G,l}^{\dag}\bm{P}_{\xi}(l\omega)\bm{Y}_{G,l},
\end{eqnarray}
where $\overline{\ \cdot\ }$ denotes the complex conjugate.
From eq. (\ref{eq. corr h}), $h_l$ can be derived likewise.

\section{Appendix C: {Derivations of the correlation functions
 $g(\theta)$ and $h(\theta)$}}
\renewcommand{\theequation}{C.\arabic{equation} }
\setcounter{equation}{0}

{For the Stuart-Landau oscillator
we used in the simulations}, we can calculate
the Fourier coefficients $\bm{Y}_{G,l}$ and $\bm{Y}_{H,l}$
as{
$\bm{Y}_{G,\pm1}=\bm{Y}_{H,\pm1}=\frac{1}{2}[1\pm i,1\mp i]^{\top}$ and
$\bm{Y}_{G,l}=\bm{Y}_{H,l}=\bm{0}\ (l\neq\pm1)$}.
Thus, from eq. (\ref{eq. corr fr}), the Fourier coefficient
$g_l$ is given by
{
$g_{\pm1} = \bm{Y}_{G,\pm l}^{\dag}\bm{Y}_{G,\pm l}
P_{\rm ex}(\omega)|_{\gamma=1}
= \frac{1}{2}\{[1+(\omega_0+3)^2]^{-1}
+[1+(\omega_0-3)^2]^{-1}\}$ and
$g_l = 0\ \ (l\neq\pm1)$},
where $\omega_0$ is a parameter.
In the same way, the Fourier coefficient
$h_l$ is given by{
$h_{\pm1} = \bm{Y}_{H,\pm l}^{\dag}\bm{Y}_{H,\pm l}
	P_{\rm ex}(\omega)|_{\omega_0=0,\gamma=3}
	= \frac{1}{2}$ and
$h_l = 0\ \ (l\neq\pm1)$}.
Substituting $g_l$ and $h_l$
to eq. (\ref{eq. corr fr}),
we can obtain {the explicit forms of $g(\theta)$ and $h(\theta)$}.


\begin{thebibliography}{10}
\expandafter\ifx\csname url\endcsname\relax\def\url#1{\texttt{#1}}\fi

\bibitem{mainen95__reliab_of_spike_timin_in_neocor_neuron}
\Name{Mainen Z.~F. \and Sejnowski T.~J.} \REVIEW{Science}{268}{1995}{1503}.

\bibitem{galan06__correl_induc_synch_of_oscil}
\Name{Gal\'an R.~F., Fourcaud-Trocm\'e N., Ermentrout G.~B. \and Urban N.~N.}
  \REVIEW{J. Neurosci.}{26}{2006}{3646}.

\bibitem{yoshida06__noise_induc_synch_of_uncoup_nonlin_system}
\Name{Yoshida K., Sato K. \and Sugamata A.} \REVIEW{J. Sound Vib.}{290}{2006}{34}.

\bibitem{utagawa08__noise_induc_synch_among_sub}
\Name{Utagawa A., Asai T., Hirose T. \and Amemiya Y.} \REVIEW{IEICE Trans.
  Fundam.}{91}{2008}{2475}.

\bibitem{zhou05__molec_commun_throug_stoch_synch}
\Name{Zhou T., Chen L. \and Aihara K.} \REVIEW{Phys. Rev. Lett.}{95}{2005}{178103}.

\bibitem{uchida04__consis_of_nonlin_system_respon}
\Name{Uchida A., McAllister R. \and Roy R.} \REVIEW{Phys. Rev. Lett.}{93}{2004}{244102}.
{
\bibitem{zhou}
\Name{Zhou C. \and Kurths J.} \REVIEW{Phys. Rev. Lett.}{88}{2002}{230602}.

\bibitem{wang}
\Name{Wang Y., Lai Y.-C., \and Zheng Z.}\REVIEW{Phys. Rev. E}{79}{2009}{056210}.
}
\bibitem{teramae04__robus_of_noise_induc_phase}
\Name{Teramae J.-N. \and Tanaka D.} \REVIEW{Phys. Rev. Lett.}{93}{2004}{204103}.

\bibitem{nakao07__noise_induc_synch_and_clust}
\Name{Nakao H., Arai K. \and Kawamura Y.} \REVIEW{Phys. Rev. Lett.}{98}{2007}{184101}.

\bibitem{yoshimura08__invar_of_frequen_differ_in}
\Name{Yoshimura K., Davis P. \and Uchida A.} \REVIEW{Prog. Theor. Phys.}{120}{2008}{621}.

\bibitem{nagai10__noise_induc_synch_of_large}
\Name{Nagai K.~H. \and Kori H.} \REVIEW{Phys. Rev. E}{81}{2010}{065202}.

\bibitem{nakao05__synch_of_limit_cycle_oscil}
\Name{Nakao H., Arai K., Nagai K., Tsubo Y. \and Kuramoto Y.} \REVIEW{Physical
  Review E}{72}{2005}{26220}.

\bibitem{yoshimura07__synch_induc_by_common_color}
\Name{Yoshimura K., Valiusaityte I. \and Davis P.} \REVIEW{Phys. Rev. E}{75}{2007}{026208}.

\bibitem{hata10__synch_of_uncoup_oscil_by}
\Name{Hata S., Shimokawa T., Arai K. \and Nakao H.} \REVIEW{Phys. Rev. E}{82}{2010}{036206}.

\bibitem{risken96__fokker_planc_equat}
\Name{Risken H.} \Book{{The Fokker-Planck equation: Methods of solution and
  applications}} (Springer Verlag) 1996.

\bibitem{cateau06__relat_between_singl_neuron_and}
\Name{C\^ateau H. \and Reyes A.~D.} \REVIEW{Phys. Rev. Lett.}{96}{2006}{058101}.

\bibitem{teramae06__noise_induc_phase_synch_of}
\Name{Teramae J. \and Tanaka D.} \REVIEW{Prog. Theor. Phys.}{161}{2006}{360}.

\bibitem{nakao10__effec_long_time_phase_dynam}
\Name{Nakao H., Teramae J.-N., Goldobin D.~S. \and Kuramoto Y.} \REVIEW{Chaos}{20}{2010}{3126}.

\bibitem{goldobin10__dynam_of_limit_cycle_oscil}
\Name{Goldobin D.~S., Teramae J.-N., Nakao H. \and Ermentrout G.~B.}
  \REVIEW{Phys. Rev. Lett.}{105}{2010}{154101}.

\bibitem{kuramoto03__chemic_oscil_waves_and_turbul}
\Name{Kuramoto Y.} \Book{{Chemical oscillations, waves, and turbulence}} (Dover
  Publications) 2003.

{
\bibitem{fhn}
\Name{FitzHugh R.}
\REVIEW{Biophys. J.}{1}{1961}{445}.

\bibitem{nagumo}
\Name{Nagumo~J., Arimoto~S. \and Yoshizawa~S.}
\REVIEW{Proc. IRE}{50}{1962}{2061}.
}

\bibitem{fellous01__frequen_depen_of_spike_timin}
\Name{Fellous J.~M., Houweling A.~R., Modi R.~H., Rao R.~P.~N., Tiesinga
  P.~H.~E. \and Sejnowski T.~J.} \REVIEW{J. Neurophysiol.}{85}{2001}{1782}.

\bibitem{hh}
\Name{Hodgkin A. \and Huxley A.} \REVIEW{J. Physiol.}{117}{1952}{500}.

\end{thebibliography}
\end{document}